\documentclass[aps,prb,twocolumn,showpacs]{revtex4}
\usepackage{graphicx}
\usepackage{amsmath}
\usepackage{amssymb}

\begin{document}

\title{Spin dynamics in the pressure-induced two-leg ladder cuprate
superconductor Sr$_{14-x}$Ca$_{x}$Cu$_{24}$O$_{41}$}

\author{Jihong Qin}

\affiliation{Department of Physics, University of Science and
Technology Beijing, Beijing 100083, China}

\author{Yu Lan}

\affiliation{Department of Physics, Jinan University, Guangzhou
510632, China}

\author{Shiping Feng}

\affiliation{Department of Physics, Beijing Normal University,
Beijing 100875, China}

\begin{abstract}
Within the two-leg $t$-$J$ ladder, the spin dynamics of the
pressure-induced two-leg ladder cuprate superconductor
Sr$_{14-x}$Ca$_{x}$Cu$_{24}$O$_{41}$ is studied based on the kinetic
energy driven superconducting mechanism. It is shown that in the
pressure-induced superconducting state, the incommensurate spin
correlation appears in the underpressure regime, while the
commensurate spin fluctuation emerges in the optimal pressure and
overpressure regimes. In particular, the spin-lattice relaxation
time is dominated by a temperature linear dependence term at low
temperature followed by a peak developed below the superconducting
transition temperature, in qualitative agreement with the
experimental observation on Sr$_{14-x}$Ca$_{x}$Cu$_{24}$O$_{41}$.
\end{abstract}


\maketitle

The doped two-leg ladder cuprate
Sr$_{14-x}$Ca$_{x}$Cu$_{24}$O$_{41}$ is a system in which a
superconducting (SC) state is realized by applying a high pressure
of $3\sim 8$GPa in the highly charge carrier doped region
\cite{Uehara96}. This pressure-induced superconductor possesses a
complex structure consisting of the Cu$_{2}$O$_{3}$ two-leg ladder
and CuO$_{2}$ chain \cite{Hiroi91,Osafune97}, then charge carriers
are transferred from CuO$_{2}$ chain unit by substituting Ca for Sr.
At the half-filling, the ground state is a spin liquid state with a
finite spin gap \cite{Eccleston98} and this gapped spin liquid state
persists even in the highly charge carrier doped region
\cite{Katano99}. Moreover, the structure of
Sr$_{14-x}$Ca$_{x}$Cu$_{24}$O$_{41}$ under high pressure remains the
same as the case in ambient pressure \cite{Isobe98}, and then the
spin background in the SC phase does not drastically alter its spin
gap properties \cite{Dagotto99}. Experimentally, by virtue of
systematic studies using the nuclear magnetic resonance (NMR) and
nuclear quadrupole resonance (NQR), the dynamical spin response of
Sr$_{14-x}$Ca$_{x}$Cu$_{24}$O$_{41}$ has been well established now
\cite{Fujiwara03,Piskunov04,Piskunov05}. In the pressure-induced SC
state, the pressure promotes the existence of low-lying spin
excitations giving rise to a residual spin susceptibility at low
temperature \cite{Piskunov04}. Furthermore, the spin-lattice
relaxation time is dominated by a temperature linear dependence term
at low temperature followed by a peak developed below the SC
transition temperature \cite{Fujiwara03}. In this case, the
interplay between the magnetic excitation and superconductivity in
two-leg ladder cuprate superconductors is of
central concern as is the case with the planar cuprate
superconductors \cite{Anderson87}.

In our earlier work \cite{He03} using the charge-spin separation
(CSS) fermion-spin theory \cite{feng04,feng07}, the dynamical spin
response of Sr$_{14-x}$Ca$_{x}$Cu$_{24}$O$_{41}$ in the {\it normal
state} has been studied, where our calculations clearly demonstrate
a crossover from the incommensurate antiferromagnetism in the weak
interchain coupling regime to commensurate spin fluctuation in the
strong interchain coupling regime. In particular, the nuclear
spin-lattice relaxation time decreases exponentially with decreasing
temperatures \cite{Katano99,Magishi98}. Furthermore, within the
kinetic energy driven SC mechanism \cite{feng0306}, we have discussed
the pressure-induced superconductivity \cite{Qin07} in
Sr$_{14-x}$Ca$_{x}$Cu$_{24}$O$_{41}$, and the result of the pressure
dependence of the SC transition temperature is in good agreement
with the corresponding experimental data of
Sr$_{14-x}$Ca$_{x}$Cu$_{24}$O$_{41}$ \cite{Isobe98}. However, in the
pressure-induced SC state, a microscopic study of the dynamical spin
response of Sr$_{14-x}$Ca$_{x}$Cu$_{24}$O$_{41}$  has not been
performed theoretically thus far although the dynamical spin response
has been measured experimentally. In this paper, we study the spin
dynamics of Sr$_{14-x}$Ca$_{x}$Cu$_{24}$O$_{41}$ in the
pressure-induced SC state within the kinetic energy driven SC
mechanism, where we calculate the dynamical spin structure factor,
and then reproduce qualitatively some main features of the
corresponding temperature dependence of the spin-lattice relaxation
time in Sr$_{14-x}$Ca$_{x}$Cu$_{24}$O$_{41}$.

The basic element of the two-leg ladder cuprates is the two-leg ladder,
which is defined as two parallel chains of ions, while the coupling
between the two chains that participates in this structure is
through rungs \cite{Hiroi91,Osafune97}. It has been shown \cite{Dagotto99} from the
experiments that the ratio of the interladder resistivity to in-ladder resistivity is
$R=\rho_{a}(T)/\rho_{c}(T)\sim 10$, this large magnitude of the resistivity anisotropy
reflects that the interladder mean free path is shorter than the interladder distance,
and the charge carriers are tightly confined to the ladders, therefore the common two-leg
ladders in the doped two-leg ladder cuprates clearly dominate the most physical properties.
In this case, it has been argued that the essential physics of the doped two-leg ladder
cuprates can be described by the $t$-$J$ ladder as \cite{Dagotto96},
\begin{eqnarray}\label{t-jmode}
H&=&-t_{\parallel}\sum_{i\hat{\eta}a\sigma}C_{ia\sigma}^{\dagger}
C_{i+\hat{\eta}a\sigma}-t_{\perp}\sum_{i\sigma}
(C_{i1\sigma}^{\dagger}C_{i2\sigma}+{\rm
H.c.})\nonumber\\
&-&\mu\sum_{ia\sigma}
C_{ia\sigma }^{\dagger}C_{ia\sigma }\nonumber\\
&+& J_{\parallel}\sum_{i\hat{\eta}a}{\bf S}_{ia}\cdot {\bf
S}_{i+\hat{\eta}a} +J_{\perp}\sum_{i}{\bf S}_{i1}\cdot{\bf S}_{i2},
\end{eqnarray}
supplemented by the local constraint $\sum_{\sigma}
C_{ia\sigma}^{\dagger}C_{ia\sigma}\leq 1$ to remove double
occupancy, where $\hat{\eta}=\pm\hat{x}$, $i$ runs over all rungs,
$\sigma(=\uparrow,\downarrow)$ and $a(=1,2)$ are spin and leg
indices, respectively, $C^{\dagger}_{ia\sigma}$ ($C_{ia\sigma}$) are
the electron creation (annihilation) operators, ${\bf S}_{ia}=
(S^{x}_{ia},S^{y}_{ia}, S^{z}_{ia})$ are the spin operators, and
$\mu$ is the chemical potential. This local constraint can be
treated properly in analytical calculations within the CSS
fermion-spin theory \cite{feng04,feng07}, $C_{ia\uparrow}=
h^{\dagger}_{ia\uparrow} S^{-}_{ia}$ and $ C_{ia\downarrow}=
h^{\dagger}_{ia\downarrow}S^{+}_{ia}$, where the spinful fermion
operator $h_{ia\sigma}= e^{-i\Phi_{i\sigma}}h_{ia}$ describes the
charge degree of freedom together with some effects of the spin
configuration rearrangements due to the presence of the doped charge
carrier itself, while the spin operator $S_{ia}$ describes the spin
degree of freedom, then the electron local constraint for the single
occupancy, $\sum_{\sigma}C^{\dagger}_{ia\sigma}C_{ia\sigma}=S^{+}_{ia}
h_{ia\uparrow}h^{\dagger}_{ia\uparrow}S^{-}_{ia}+S^{-}_{ia}h_{ia\downarrow}
h^{\dagger}_{ia\downarrow}S^{+}_{ia}=h_{ia}h^{\dagger}_{ia}(S^{+}_{ia}
S^{-}_{ia}+S^{-}_{ia}S^{+}_{ia})=1-h^{\dagger}_{ia}h_{ia}\leq 1$, is
satisfied in analytical calculations. Although in common sense $h_{ia\sigma}$
is not a real spinful fermion operator, it behaves like a spinful fermion.
This is followed from a fact that the spinless fermion $h_{ia}$ and spin
operators $S^{+}_{ia}$ and $S^{-}_{ia}$ obey the anticommutation relation and
Pauli spin algebra, respectively, it is then easy to show that the spinful
fermion $h_{ia\sigma}$ also obey the same anticommutation relation as the
spinless fermion $h_{ia}$. In particular, it has been shown that under the
decoupling scheme, this CSS fermion-spin representation is a natural
representation of the constrained electron defined in the restricted Hilbert
space without double electron occupancy \cite{feng07}. Moreover, these charge
carrier and spin are gauge invariant \cite{feng04,feng07}, and in this sense,
they are real and can be interpreted as the physical excitations
\cite{laughlin97}. In this CSS fermion-spin representation, the low-energy
behavior of the $t$-$J$ ladder Hamiltonian (1) can be expressed as,
\begin{eqnarray}\label{t-jmode1}
H&=&t_{\parallel}\sum_{i\hat{\eta}a}
(h^{\dagger}_{i+\hat{\eta}a\uparrow} h_{ia\uparrow}S^{+}_{ia}
S^{-}_{i+\hat{\eta}a}+ h^{\dagger}_{i+\hat{\eta}a\downarrow}
h_{ia\downarrow}S^{-}_{ia}S^{+}_{i+\hat{\eta}a})\nonumber\\
&+&t_{\perp}\sum_{i} (h^{\dagger}_{i2\uparrow}h_{i1\uparrow}
S^{+}_{i1} S^{-}_{i2} +h^{\dagger}_{i1\uparrow}h_{i2\uparrow}
S^{+}_{i2}S^{-}_{i1}\nonumber\\
&+&h^{\dagger}_{i2\downarrow}h_{i1\downarrow} S^{-}_{i1}S^{+}_{i2}
+h^{\dagger}_{i1\downarrow}h_{i2\downarrow} S^{-}_{i2}S^{+}_{i1})\nonumber\\
&+&\mu\sum_{ia\sigma}h^{\dagger}_{ia\sigma}h_{ia\sigma}\nonumber\\
&+&{J_{\parallel\rm eff}}\sum_{i\hat{\eta}a} {\bf S}_{ia}\cdot {\bf
S}_{i+\hat{\eta}a}+{J_{\perp\rm eff}}\sum_{i}{\bf S}_{i1} \cdot {\bf
S}_{i2},
\end{eqnarray}
where $J_{\parallel\rm eff}=J_{\parallel}(1-p)^{2}$, $J_{\perp\rm eff}=J_{\perp}(1-p)^{2}$,
and $p=\langle h^{\dagger}_{ia\sigma}h_{ia\sigma}\rangle=\langle h^{\dagger}_{ia}h_{ia}\rangle$
is the charge carrier doping concentration. Although the CSS fermion-spin representation is a
natural representation for the constrained electron under the decoupling scheme \cite{feng07},
so long as $h^{\dagger}_{ia}h_{ia}=1$, $\sum_{\sigma}C^{\dagger}_{ia\sigma}C_{ia\sigma}=0$, no
matter what the values of $S^{+}_{ia}S^{-}_{ia}$ and $S^{-}_{ia}S^{+}_{ia}$ are, therefore it
means that a {\it spin} even to an empty site has been assigned. Obviously, this insignificant
defect is originated from the decoupling approximation. It has been shown \cite{feng94} that
this defect can be cured by introducing a projection operator $P_{i}$, i.e., the electron operator
$C_{ia\sigma}$ with the single occupancy local constraint can be mapped exactly using the CSS
fermion-spin transformation defined with an additional projection operator $P_{i}$. However,
this projection operator is cumbersome to handle in the many cases, and it has been dropped in
the actual calculations \cite{He03,feng04,feng07,feng0306,Qin07}. It has been shown
\cite{feng04,feng07,feng94,plakida02} that such treatment leads to errors of the order $p$ in
counting the number of spin states, which is negligible for small dopings. Moreover, the electron
single occupancy local constraint still is exactly obeyed even in the mean-field (MF)
approximation. These are why the theoretical results \cite{He03,Qin07} obtained from the $t$-$J$
ladder model (\ref{t-jmode1}) based on the CSS fermion-spin theory are in qualitative agreement
with the experimental observation on the doped two-leg ladder cuprates.

It has been shown from the experiments \cite{Uehara96,Isobe98,Nagata98} that the pressure-induced
SC state in the doped two-leg ladder cuprate Sr$_{14-x}$Ca$_{x}$Cu$_{24}$O$_{41}$ is also
characterized by the electron Cooper pairs as in the conventional superconductors
\cite{Schrieffer64}, forming SC quasiparticles. However, because there are two coupled $t$-$J$
chains in the pressure-induced two-leg ladder cuprate superconductors, the energy spectrum has two
branches, and therefore the one-particle spin Green's function, the charge carrier normal and
anomalous Green's functions are matrices, and can be expressed as,
$D(i-j,\tau-\tau^{\prime})=D_{L}(i-j,\tau-\tau^{\prime})+\sigma_{x}D_{T}(i-j,\tau-\tau^{\prime})$,
$g(i-j,\tau-\tau^{\prime})=g_{L}(i-j,\tau-\tau^{\prime})+\sigma_{x}g_{T}(i-j,\tau-\tau^{\prime})$,
$\Gamma^{\dagger}(i-j,\tau-\tau^{\prime})=\Gamma^{\dagger}_{L}(i-j,\tau-\tau^{\prime})+\sigma_{x}
\Gamma^{\dagger}_{T}(i-j,\tau-\tau^{\prime})$, respectively, where the corresponding longitudinal
and transverse parts are defined as
$D_{L}(i-j,\tau-\tau^{\prime})=-\langle T_{\tau}S_{ia}^{+}(\tau)S_{ja}^{-}(\tau^{\prime})\rangle$,
$g_{L}(i-j,\tau-\tau^{\prime})=-\langle T_{\tau}h_{ia\sigma}(\tau)h_{ja\sigma}^{\dagger}(\tau^{\prime})
\rangle$, $\Gamma^{\dagger}_{L}(i-j,\tau-\tau^{\prime})=-\langle T_{\tau}h_{ia\uparrow}(\tau)
h_{ja\downarrow}^{\dagger}(\tau^{\prime})\rangle$, and $D_{T}(i-j,\tau-\tau^{\prime})=-\langle
T_{\tau}S_{ia}^{+}(\tau)S_{ja^{\prime}}^{-}(\tau^{\prime})\rangle$, $g_{T}(i-j,\tau-\tau^{\prime})=
-\langle T_{\tau}h_{ia\sigma}(\tau)h_{ja^{\prime}\sigma}^{\dagger}(\tau^{\prime})\rangle$,
$\Gamma^{\dagger}_{T}(i-j,\tau-\tau^{\prime})=-\langle T_{\tau}h_{ia\uparrow}(\tau)
h_{ja^{\prime}\downarrow}^{\dagger}(\tau^{\prime})\rangle$, with $a^{\prime}\neq a$. In this case,
the order parameters for the electron Cooper pair also is a matrix $\Delta=\Delta_{L}+\sigma_{x}
\Delta_{T}$, with the longitudinal and transverse SC order parameters are defined as,
\begin{subequations}
\begin{eqnarray}
\Delta_{L}&=&\langle
C^{\dagger}_{ia\uparrow}C^{\dagger}_{i+\hat{\eta}a\downarrow}
-C^{\dagger}_{ia\downarrow}C^{\dagger}_{i+\hat{\eta}a\uparrow}\rangle
=\langle
h_{ia\uparrow}h_{i+\hat{\eta}a\downarrow}S^{+}_{ia}S^{-}_{i+\hat{\eta}a}\nonumber\\
&-&h_{ia\downarrow}
h_{i+\hat{\eta}a\uparrow}S^{-}_{ia}S^{+}_{i+\hat{\eta}a}\rangle
=-\chi_{\parallel}\Delta_{hL},\\
\label{gap1} \Delta_{T}&=&\langle C^{\dagger}_{i1\uparrow}
C^{\dagger}_{i2\downarrow}- C^{\dagger}_{i1\downarrow}
C^{\dagger}_{i2\uparrow}\rangle=\langle h_{i1\uparrow}
h_{i2\downarrow}S^{+}_{i1}S^{-}_{i2}\nonumber\\
&-&h_{i1\downarrow}h_{i2\uparrow}S^{-}_{i1}S^{+}_{i2}
\rangle=-\chi_{\perp} \Delta_{hT}, \label{gap2}
\end{eqnarray}
\end{subequations}
respectively, where the spin correlation functions
$\chi_{\parallel}=\langle S_{ia}^{+}S_{i+\hat{\eta}a}^{-}\rangle$
and $\chi_{\perp}=\langle S^{+}_{i1}S^{-}_{i2}\rangle$, and the
longitudinal and transverse charge carrier pairing order parameters
are expressed as $\Delta_{hL}=\langle h_{ja\downarrow}h_{ia\uparrow}
-h_{ja\uparrow}h_{ia\downarrow}\rangle$ and $\Delta_{hT}=\langle
h_{i2\downarrow}h_{i1\uparrow}-h_{i2\uparrow}h_{i1\downarrow}
\rangle$, respectively.

\begin{figure}[h!]
\center\includegraphics[scale=0.5]{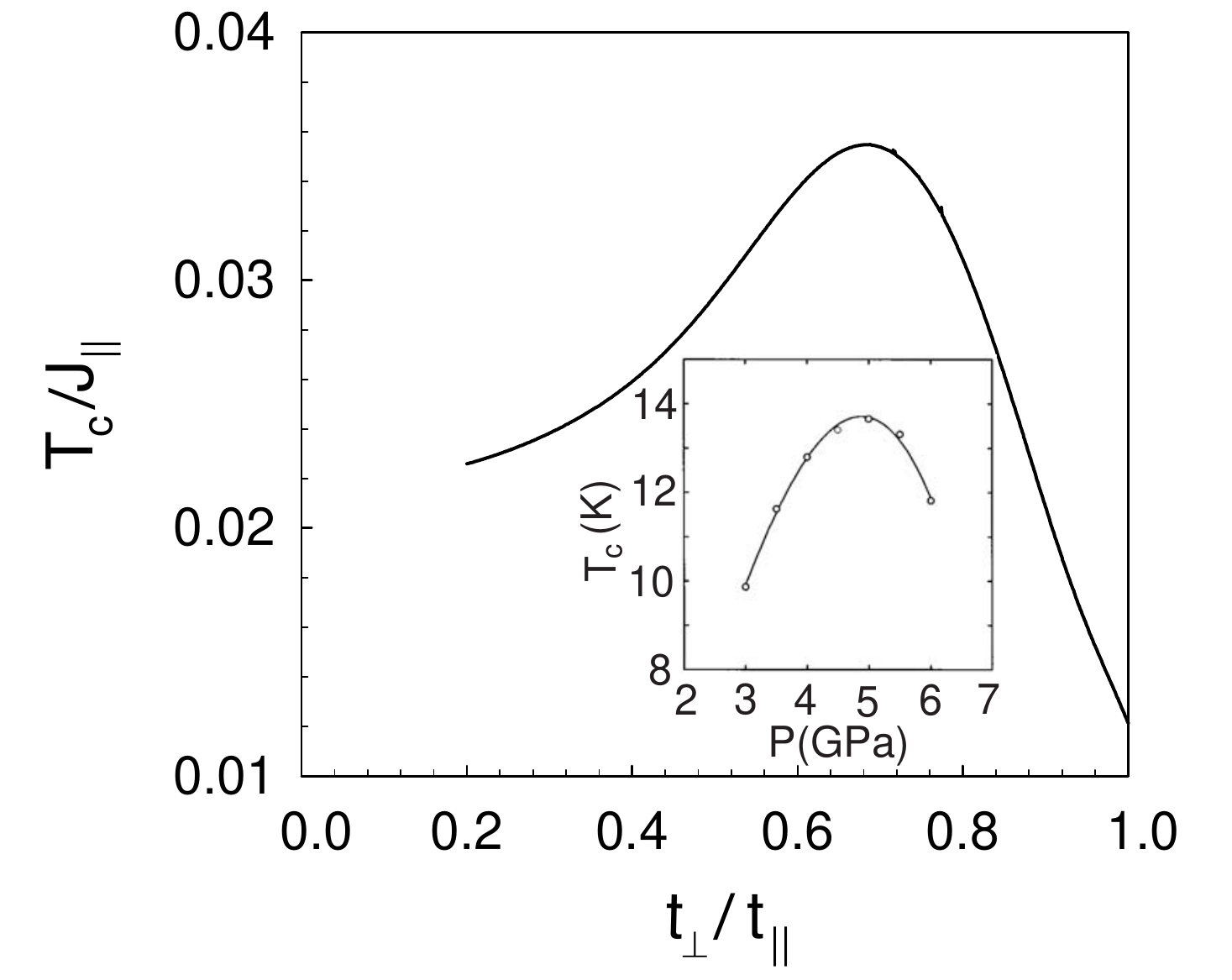} \caption{The
superconducting transition temperature as a function of
$t_{\perp}/t_{\parallel}$ for $t_{\parallel}/J_{\parallel}=2.5$ at
$p=0.25$. Inset: the experimental result of
Sr$_{14-x}$Ca$_{x}$Cu$_{24}$O$_{41}$ taken from Ref. \cite{Isobe98}.
\label{fig1}}
\end{figure}

At ambient pressure, the exchange coupling $J_{\parallel}$ along the
legs is greater than exchange coupling $J_{\perp}$ across a rung,
i.e., $J_{\parallel}>J_{\perp}$, and similarly the hopping
$t_{\parallel}$ along the legs is greater than the rung hopping
strength $t_{\perp}$, i.e., $t_{\parallel}>t_{\perp}$. In this case,
the doped two-leg ladder cuprate Sr$_{14-x}$Ca$_{x}$Cu$_{24}$O$_{41}$
is highly anisotropic material \cite{Eccleston98,Dagotto99,Magishi98}.
However, pressure for realizing superconductivity in doped two-leg
ladder cuprates plays a role of stabilizing the metallic state and
suppressing anisotropy within the ladders. This is followed an
experimental fact
\cite{Uehara96,Isobe98,Fujiwara03,Nagata98,Ohta97,Kato96} that the
distance between ladders and chains is reduced with increasing
pressure, and then the coupling between ladders and chains is
enhanced. This leads to that the values of $J_{\perp}/J_{\parallel}$
and $t_{\perp}/t_{\parallel}$ increase with increasing pressure. In
other words, the pressurization induces anisotropy shrinkage on
doped two-leg ladder cuprates, and then there is a tendency toward
the isotropy for doped two-leg ladders
\cite{Uehara96,Isobe98,Fujiwara03,Nagata98,Ohta97,Kato96}. These
experimental results explicitly imply that the values of
$J_{\perp}/J_{\parallel}$ and $t_{\perp}/t_{\parallel}$ of doped
two-leg ladder cuprates are closely related to the pressurization,
and therefore the pressure effects can be imitated by a variation of
the values of $J_{\perp}/J_{\parallel}$ and $t_{\perp}/t_{\parallel}$.
On the other hand, as we have mentioned above, the structure of
Sr$_{14-x}$Ca$_{x}$Cu$_{24}$O$_{41}$ by applying a high pressure of $3\sim 8$GPa
remains the same as the case in ambient pressure \cite{Isobe98}, and then the
spin background in the SC phase does not drastically alter its spin
gap properties \cite{Dagotto99}. In this case, the pressure-induced
superconductivity in Sr$_{14-x}$Ca$_{x}$Cu$_{24}$O$_{41}$ has been
discussed \cite{Qin07} within the kinetic energy driven SC mechanism
\cite{feng0306}, and a dome-shaped SC transition temperature $T_{c}$
versus pressure curve is obtained, where the variation of
$(t_{\perp}/ t_{\parallel})^{2}$ under the pressure is chosen the
same as that of $J_{\perp}/J_{\parallel}$, i.e.,
$(t_{\perp}/t_{\parallel})^{2} =J_{\perp}/J_{\parallel}$. For the
convenience in the following discussions, this result of the
dome-shaped SC transition temperature $T_{c}$ versus pressure is
replotted in Fig. \ref{fig1} in comparison with the corresponding
experimental result \cite{Isobe98} of
Sr$_{14-x}$Ca$_{x}$Cu$_{24}$O$_{41}$ (inset), where the maximal SC
transition temperature occurs around the optimal pressure
($t_{\perp}/t_{\parallel}\approx 0.7$), then decreases in both
underpressure ( $t_{\perp}/t_{\parallel}<0.7$) and overpressure
($t_{\perp}/t_{\parallel}> 0.7$) regimes, and is in good agreement
with the corresponding experimental data of
Sr$_{14-x}$Ca$_{x}$Cu$_{24}$O$_{41}$ \cite{Isobe98}.

Following our previous discussions \cite{Qin07,Qin06}, the
longitudinal and transverse parts of the charge carrier normal and
anomalous Green functions of the pressure-induced two-leg ladder
cuprate superconductors can be obtained as,
\begin{widetext}
\begin{subequations}\label{holegreenfuction}
\begin{eqnarray}
g_{L}(k,\omega)&=&{1\over 2}\sum_{\nu=1,2}Z^{(\nu)}_{FA}\left
({U^{2}_{\nu k}\over\omega-E_{\nu k}}+
{V^{2}_{\nu k}\over \omega+E_{\nu k}}\right), \\
g_{T}(k,\omega)&=&{1\over 2}\sum_{\nu=1,2}(-1)^{\nu+1}
Z^{(\nu)}_{FA }\left ({U^{2}_{\nu k}\over\omega-E_{\nu k}}
+{V^{2}_{\nu k}\over\omega+E_{\nu k}}\right ), \\
\Gamma^{\dagger}_{L}(k,\omega)&=&-{1\over 2}\sum_{\nu=1,2}
Z^{(\nu)}_{FA }{\bar{\Delta}_{hz}^{(\nu)}(k)\over 2E_{\nu k}}
\left ({1\over \omega-E_{\nu k}}-{1\over\omega+E_{\nu k}}\right ), \\
\Gamma^{\dagger}_{T}(k,\omega)&=&-{1\over 2}\sum_{\nu=1,2}
(-1)^{\nu+1}Z^{(\nu)}_{FA }{\bar{\Delta}_{hz}^{(\nu)}(k)\over
2E_{\nu k}} \left ({1\over\omega-E_{\nu k}}-{1\over\omega+E_{\nu k}}\right ),
\end{eqnarray}
\end{subequations}
\end{widetext}
where $Z^{(1)-1}_{FA}=Z^{-1}_{F1}-Z^{-1}_{F2}$ and $Z^{(2)-1}_{FA}
=Z^{-1}_{F1}+Z^{-1}_{F2}$ with the charge carrier longitudinal and
transverse quasiparticle coherent weights $Z_{F1}$ and $Z_{F2}$,
respectively, the charge carrier quasiparticle coherence factors
$U^{2}_{\nu k}=[1+\bar{\xi}_{\nu k}/E_{\nu k}]/2$ and $V^{2}_{\nu
k}=[1-\bar{\xi}_{\nu k}/E_{\nu k}]/2$, the renormalized charge
carrier excitation spectrum ${\bar\xi}_{\nu
k}=Z^{(\nu)}_{FA}\xi_{\nu k}$, with the MF charge carrier excitation
spectrum $\xi_{\nu k}
=2t_{\parallel}\chi_{\parallel}\cos{k}+\mu+\chi_{\perp} t_{\perp}
(-1)^{\nu+1}$, the renormalized charge carrier pair gap function
$\bar{\Delta}_{hz}^{(\nu)}(k)=Z^{(\nu)}_{FA}[2\bar{\Delta}_{hL}\cos{k}+
(-1)^{\nu+1}\bar{\Delta}_{hT}]$, and the charge carrier
quasiparticle dispersion $E_{\nu k}=\sqrt{[{\bar\xi}_{\nu k}]^{2}+
\mid\bar{\Delta}_{hz}^{(\nu)}(k)\mid^{2}}$. While the longitudinal
and transverse parts of the MF spin Green's function
$D^{(0)}(k,\omega)$ can be obtained as \cite{He03},
\begin{subequations}\label{MFGreen}
\begin{eqnarray}
D^{(0)}_{L}(k,\omega)&=&{1\over 2}\sum_{\mu=1,2}{B_{\mu k}\over \omega^{2}-\omega^{2}_{\mu k}}, \\
D^{(0)}_{T}(k,\omega)&=&{1\over 2}\sum_{\mu=1,2}(-1)^{\mu+1}{B_{\mu k}\over \omega^{2}-\omega^{2}_{\mu k}},
\end{eqnarray}
\end{subequations}
where $B_{\mu k}=\lambda [A_{1}{\rm cos}k- A_{2}]-J_{\perp\rm
eff}[\chi_{\perp}+
2\chi^{z}_{\perp}(-1)^{\mu}][\epsilon_{\perp}+(-1)^{\mu}]$,
$\lambda=4J_{\parallel\rm eff}$,
$A_{1}=2\epsilon_{\parallel}\chi^{z}_{\parallel}+ \chi_{\parallel}$,
$A_{2}=\epsilon_{\parallel}\chi_{\parallel}+ 2\chi^{z}_{\parallel}$,
$\epsilon_{\parallel}=1+2t_{\parallel}\phi_{\parallel}/J_{\parallel\rm
eff}$, and $\epsilon_{\perp}=1+ 4t_{\perp} \phi_{\perp}/J_{\perp\rm
eff}$, with the spin correlation functions
$\chi^{z}_{\parallel}=\langle
S_{ia}^{z}S_{i+\hat{\eta}a}^{z}\rangle$, $\chi^{z}_{\perp}=\langle
S_{i1}^{z} S_{i2}^{z}\rangle$, the charge carrier particle-hole
order parameters $\phi_{\parallel}=\langle
h^{\dagger}_{ia\sigma}h_{i+\hat{\eta}a\sigma}\rangle$,
$\phi_{\perp}=\langle h^{\dagger}_{i1\sigma}h_{i2\sigma}\rangle$.
The MF spin excitation spectra, $\omega^{2}_{\mu{\bf
k}}=\alpha\epsilon_{\parallel}\lambda^{2}A_{1}{\rm cos}^{2}k/2
+[X_{1}+X_{2}(-1)^{\mu+1}] {\rm cos}k+X_{3}+X_{4}(-1)^{\mu+1}$,
where $X_{1}=-\epsilon_{\parallel}\lambda^{2}[(\alpha
A_{2}+2A_{4})/4+A_{3}]-\alpha\lambda J_{\perp\rm eff}
[\epsilon_{\parallel}(C^{z}_{\perp}+\chi^{z}_{\perp})+
\epsilon_{\perp}(C_{\perp}+\epsilon_{\parallel}\chi_{\perp})/2]$,
$X_{2}=\alpha\lambda J_{\perp\rm eff}[(\epsilon_{\perp}
\chi_{\parallel}+\epsilon_{\parallel}\chi_{\perp})/2+
\epsilon_{\parallel}\epsilon_{\perp}(\chi^{z}_{\perp}+
\chi^{z}_{\parallel})]$, $X_{3}=\lambda^{2}[A_{3}-\alpha
\epsilon_{\parallel}A_{1}/4+\epsilon^{2}_{\parallel}A_{4}/2]+
\alpha\lambda J_{\perp\rm eff}[\epsilon_{\parallel}\epsilon_{\perp}
C_{\perp}+ 2C^{z}_{\perp}]+J^{2}_{\perp\rm eff}
(\epsilon^{2}_{\perp}+1)/4$, $X_{4}=-\alpha\lambda J_{\perp\rm eff}
[\epsilon_{\parallel}\epsilon_{\perp}\chi_{\parallel}/2+
\epsilon_{\perp}(\chi^{z}_{\parallel}+C^{z}_{\perp})+
\epsilon_{\parallel}C_{\perp}/2]-\epsilon_{\perp}J^{2}_{\perp\rm
eff} /2$, with $A_{3}=\alpha C^{z}_{\parallel}+(1-\alpha)/8$,
$A_{4}=\alpha C_{\parallel}+(1-\alpha)/4$, and the spin correlation
functions $C_{\parallel}=\sum_{\hat{\eta}\hat{\eta'}}\langle
S_{i+\hat{\eta}a}^{+} S_{i+\hat{\eta'}a}^{-}\rangle/4$,
$C^{z}_{\parallel}= \sum_{\hat{\eta}\hat{\eta'}}\langle
S_{i+\hat{\eta}a}^{z} S_{i+\hat{\eta'}a}^{z}\rangle/4$,
$C_{\perp}=\sum_{\hat{\eta}} \langle S_{i2}^{+}
S_{i+\hat{\eta}1}^{-}\rangle/2$, and $C^{z}_{\perp}=
\sum_{\hat{\eta}}\langle S_{i1}^{z}S_{i+ \hat{\eta}2}^{z}\rangle/2$.
In order to satisfy the sum rule for the correlation function
$\langle S^{+}_{ia}S^{-}_{ia}\rangle =1/2$ in the absence of the
antiferromagnetic (AF) long-range-order, a decoupling parameter
$\alpha$ has been introduced in the MF calculation, which can be
regarded as the vertex correction \cite{He03,Qin07}, then all these
MF order parameters, decoupling parameter, chemical potential,
charge carrier longitudinal and transverse quasiparticle coherent
weights $Z_{F1}$ and $Z_{F2}$, and longitudinal and transverse
charge carrier pair gap parameters $\bar{\Delta}_{hL}$ and
$\bar{\Delta}_{hT}$ are determined by the self-consistent
calculation \cite{Qin07,Qin06}.

In the CSS fermion-spin theory \cite{feng04,feng07}, the
AF fluctuation is dominated by the scattering of
the spins. In the normal state, the spins move in the charge carrier
background, therefore the spin self-energy (then full spin Green's
function) in the normal state has been obtained in terms of the
collective mode in the charge carrier particle-hole channel
\cite{He03}. With the help of this full spin Green's function in the
normal state, the dynamical spin response of
Sr$_{14-x}$Ca$_{x}$Cu$_{24}$O$_{41}$ in the normal state has been
discussed \cite{He03}, and the results are consistent with the
corresponding experimental results \cite{Katano99,Magishi98}.
However, in the present pressure-induced SC state, the spins move in
the charge carrier pairing background. On the other hand, we \cite{qin02} have
discussed the optical and transport properties of the doped two-leg ladder cuprates
in the underdoped and optimally doped regimes by considering the fluctuations
around the MF solution, where the dominant dynamical effect is due to the strong interaction
between the charge carriers and spins in the $t$-$J$ ladder Hamiltonian (\ref{t-jmode1}).
We believe that this strong interaction between the charge carriers and spins also will
dominate the spin dynamics within the same doping regimes. In this case, we can
calculate the spin self-energy (then the full spin Green's function)
in terms of the collective modes in the charge carrier particle-hole
and particle-particle channels. Following our previous discussions
for the normal-state case \cite{He03}, the full spin Green's
function in the present pressure-induced SC state can be obtained as,
\begin{eqnarray}\label{SGF1}
D(k,\omega)=D^{(0)}(k,\omega)+D^{(0)}(k,\omega)\Sigma^{(s)}(k,\omega)D(k,\omega),
\end{eqnarray}
then the dynamical spin structure factor of the pressure-induced two-leg
ladder cuprate superconductors can be obtained explicitly in terms of full spin
Green's function (\ref{SGF1}) as,
\begin{widetext}
\begin{eqnarray}\label{DSSF}
S(k,\omega)&=& -2[1+n_{B}(\omega)][{\rm Im}D_{L}(k,\omega)+{\rm
Im}D_{T}(k,\omega)]
\nonumber \\
&=& -2[1+n_{B}(\omega)]{B_{1k}^{2}{\rm
Im}\Sigma^{(1)}_{s}(k,\omega)\over [\omega^{2}-
(\omega_{1k})^{2}-B_{1k}{\rm
Re}\Sigma^{(1)}_{s}(k,\omega)]^{2}+[B_{1{\bf k}}{\rm Im}
\Sigma^{(1)}_{s}(k,\omega)]^{2}},
\end{eqnarray}
\end{widetext}
where $n_{B}(\omega)$ is the boson distribution function, ${\rm Im}
\Sigma^{(1)}_{s}(k,\omega)={\rm Im}\Sigma^{(s)}_{L}(k,\omega)+{\rm
Im} \Sigma^{(s)}_{T}(k,\omega)$ and ${\rm Re}
\Sigma^{(1)}_{s}(k,\omega)={\rm Re} \Sigma^{(s)}_{L}(k, \omega)+{\rm
Re}\Sigma^{(s)}_{T}(k,\omega)$, while ${\rm Im}
\Sigma^{(s)}_{L}(k,\omega) [{\rm Re}\Sigma^{(s)}_{L}(k, \omega)]$
and ${\rm Im}\Sigma^{(s)}_{T} (k,\omega) [{\rm Re}
\Sigma^{(s)}_{T}(k,\omega)]$ are the corresponding imaginary (real)
parts of the spin longitudinal and transverse self-energy,
respectively, and this spin self-energy $\Sigma^{(s)}(k,
\omega)=\Sigma^{(s)}_{L}(k,\omega)+ \sigma_{x}
\Sigma^{(s)}_{T}(k,\omega)$ with the longitudinal and transverse
parts can be evaluated in terms of the charge carrier normal and
anomalous Green functions in Eq. (4) and MF spin Green's function
$D^{(0)}(k,\omega)$ in Eq. (5) as,
\begin{widetext}
\begin{subequations}\label{SSE2}
\begin{eqnarray}
\Sigma^{(s)}_{L}(k,\omega)&=& {1\over 32N^{2}}\sum_{p,q}\sum_{\mu\nu\nu'}\Pi_{\mu\nu\nu'}(p,q,k),\\
\Sigma^{(s)}_{T}(k,\omega)&=& {1\over 32N^{2}}\sum_{p,q}\sum_{\mu\nu\nu'}(-1)^{\mu+\nu+\nu'+1}
\Pi_{\mu\nu\nu'}(p,q,k),
\end{eqnarray}
\end{subequations}
\end{widetext}
where the kernel function,
\begin{widetext}
\begin{eqnarray}
\Pi_{\mu\nu\nu'}(p,q,k)&=&[C_{\mu\nu'}(p-k)+C_{\mu\nu}(p+q+k)]{B_{\mu
k+q}\over
\omega_{\mu k+q}}Z^{(\nu)}_{FA}Z^{(\nu')}_{FA}\nonumber \\
&\times&\left({K^{(1)}_{\mu\nu\nu'}(p,q,k)\over
\omega^{2}-(\omega_{\mu k+q}+E_{\nu p}-E_{\nu'
p+q})^{2}}+{K^{(2)}_{\mu\nu\nu'}(p,q,k)\over
\omega^{2}-(\omega_{\mu k+q}-E_{\nu p}+E_{\nu' p+q})^{2}}\right.\nonumber\\
&+&\left. {K^{(3)}_{\mu\nu\nu'}(p,q,k)\over\omega^{2}-(\omega_{\mu
k+q}+E_{\nu p} +E_{\nu'
p+q})^{2}}+{K^{(4)}_{\mu\nu\nu'}(p,q,k)\over\omega^{2}-(\omega_{\mu
k+q}-E_{\nu p}- E_{\nu' p+q})^{2}}\right),~~~~
\end{eqnarray}
\end{widetext}
with
$C_{\mu\nu}(k)=[2t_{\parallel}{\cos}k+(-1)^{\mu+\nu}t_{\perp}]^{2}$,
and
\begin{widetext}
\begin{subequations}
\begin{eqnarray}
K^{(1)}_{\mu\nu\nu'}(p,q,k)&=&\left
({\bar{\Delta}^{(\nu)}_{hz}(p)\over E_{\nu p}}
{\bar{\Delta}^{(\nu')}_{hz}(p+q)\over E_{\nu'
p+q}}-1-{\bar{\xi}_{\nu p}\over E_{\nu p}} {\bar{\xi}_{\nu'
p+q}\over E_{\nu' p+q}}\right)(\omega_{\mu k+q}+E_{\nu p}-E_{\nu'
p+q})
\nonumber \\
&\times& \{n_{B}(\omega_{\mu k+q})[n_{F}(E_{\nu p})-n_{F}(E_{\nu'
p+q})]-n_{F}(-E_{\nu p})
n_{F}(E_{\nu' p+q})\},\\
K^{(2)}_{\mu\nu\nu'}(p,q,k)&=&\left
({\bar{\Delta}^{(\nu)}_{hz}(p)\over E_{\nu p}}
{\bar{\Delta}^{(\nu')}_{hz}(p+q)\over E_{\nu'
p+q}}-1-{\bar{\xi}_{\nu p}\over E_{\nu p}} {\bar{\xi}_{\nu'
p+q}\over E_{\nu' p+q}}\right )(\omega_{\mu k+q}-E_{\nu p}+E_{\nu'
p+q})
\nonumber \\
&\times&\{n_{B}(\omega_{\mu k+q})[n_{F}(E_{\nu' p+q})-n_{F}(E_{\nu
p})]-n_{F}(E_{\nu p})
n_{F}(-E_{\nu' p+q})\},\\
K^{(3)}_{\mu\nu\nu'}(p,q,k)&=&\left
({\bar{\Delta}^{(\nu)}_{hz}(p)\over E_{\nu p}}
{\bar{\Delta}^{(\nu')}_{hz}(p+q)\over E_{\nu'
p+q}}+1-{\bar{\xi}_{\nu p}\over E_{\nu p}} {\bar{\xi}_{\nu'
p+q}\over E_{\nu' p+q}}\right) (\omega_{\mu k+q}+E_{\nu p}+E_{\nu'
p+q})
\nonumber\\
&\times&\{n_{B}(\omega_{\mu k+q})[n_{F}(-E_{\nu p})-n_{F}(E_{\nu'
p+q})]+n_{F}(-E_{\nu p})
n_{F}(-E_{\nu' p+q})\},~~~~~\\
K^{(4)}_{\mu\nu\nu'}(p,q,k)&=&\left
({\bar{\Delta}^{(\nu)}_{hz}(p)\over E_{\nu p}}
{\bar{\Delta}^{(\nu')}_{hz}(p+q)\over E_{\nu'
p+q}}+1-{\bar{\xi}_{\nu p}\over E_{\nu p}} {\bar{\xi}_{\nu'
p+q}\over E_{\nu' p+q}}\right) (\omega_{\mu k+q}-E_{\nu p}-E_{\nu'
p+q})
\nonumber\\
&\times&\{n_{B}(\omega_{\mu k+q})[n_{F}(E_{\nu p})+n_{F}(E_{\nu'
p+q})-1]+n_{F}(E_{\nu p}) n_{F}(E_{\nu' p+q})\},
\end{eqnarray}
\end{subequations}
\end{widetext}
where $n_{F}(\omega)$ is the fermion distribution function.

\begin{figure}[h!]
\center\includegraphics[scale=0.6]{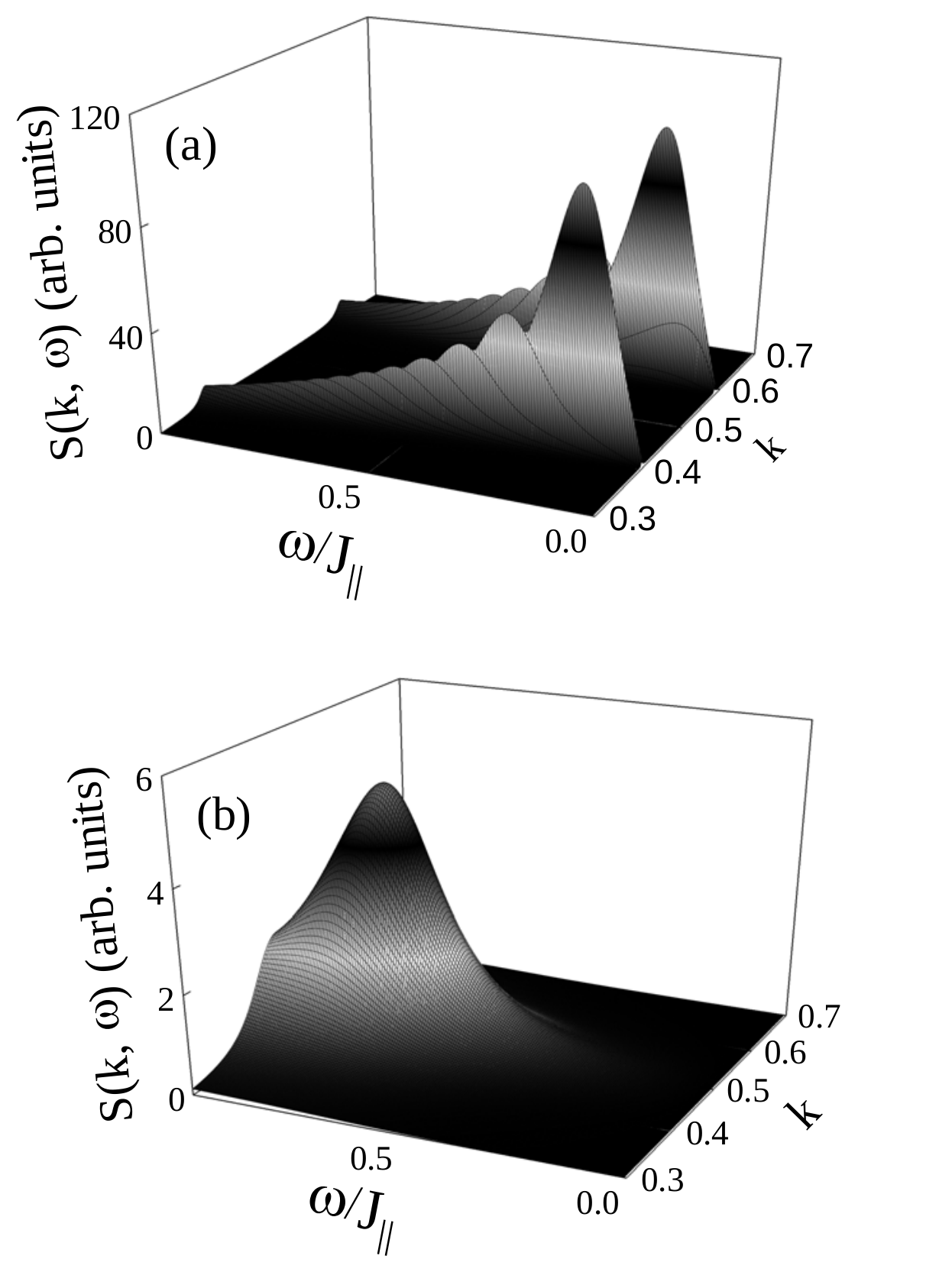} \caption{The dynamical
spin structure factor in the ($k,\omega$) plane at $p=0.20$ with
$T=0$ for $t_{\parallel}/J_{\parallel}=2.5$ and (a)
$t_{\perp}/t_{\parallel}= 0.4$ and (b)
$t_{\perp}/t_{\parallel}=0.8$. \label{fig2}}
\end{figure}

\begin{figure}[h!]
\center\includegraphics[scale=0.5]{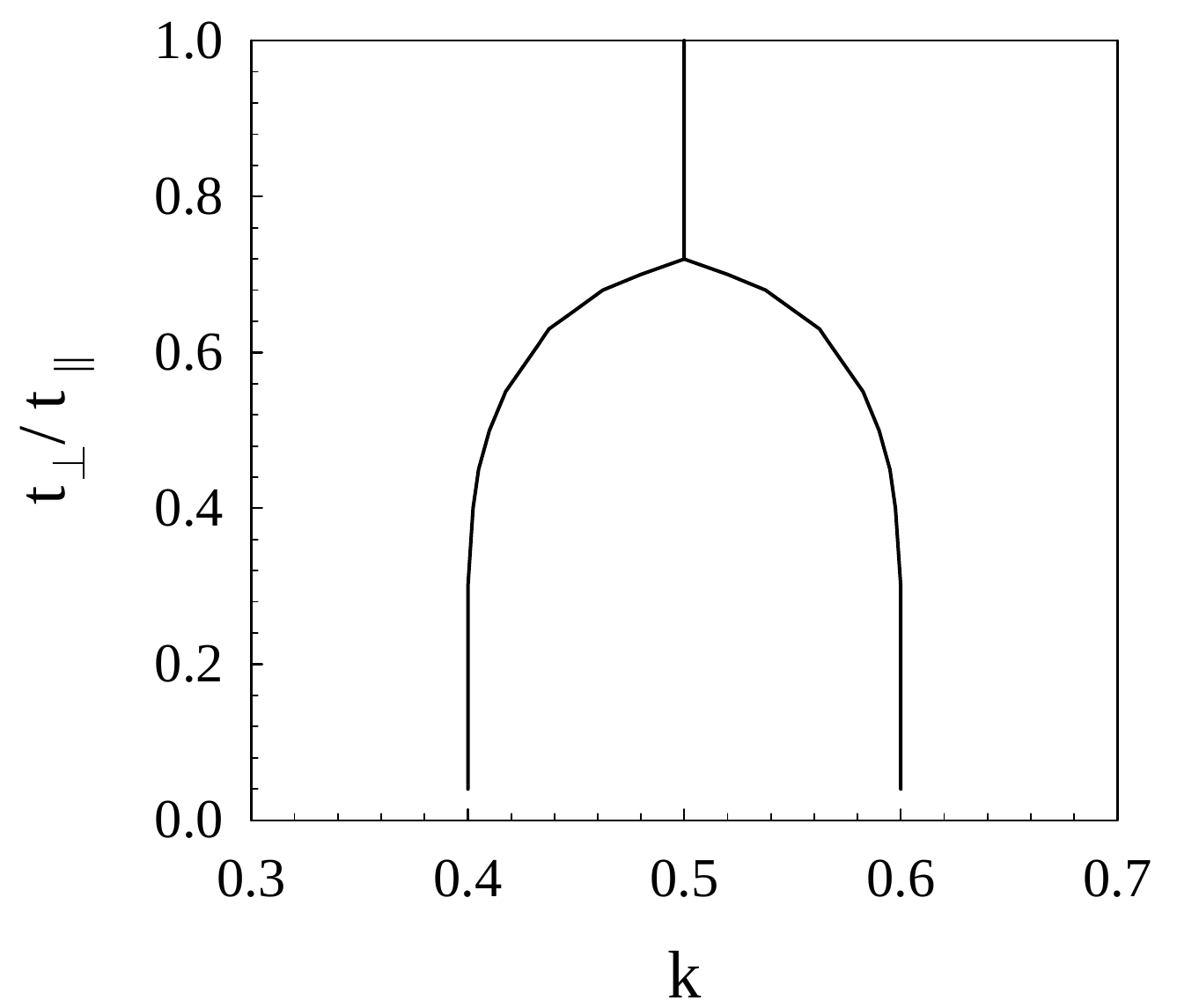} \caption{The position
of the magnetic scattering peaks as a function of
$t_{\perp}/t_{\parallel}$ at $p=0.20$ with $T=0$ for
$t_{\parallel}/J_{\parallel}=2.5$ and $\omega=0.4J_{\parallel}$.
\label{fig3}}
\end{figure}

\begin{figure}[h!]
\center\includegraphics[scale=0.5]{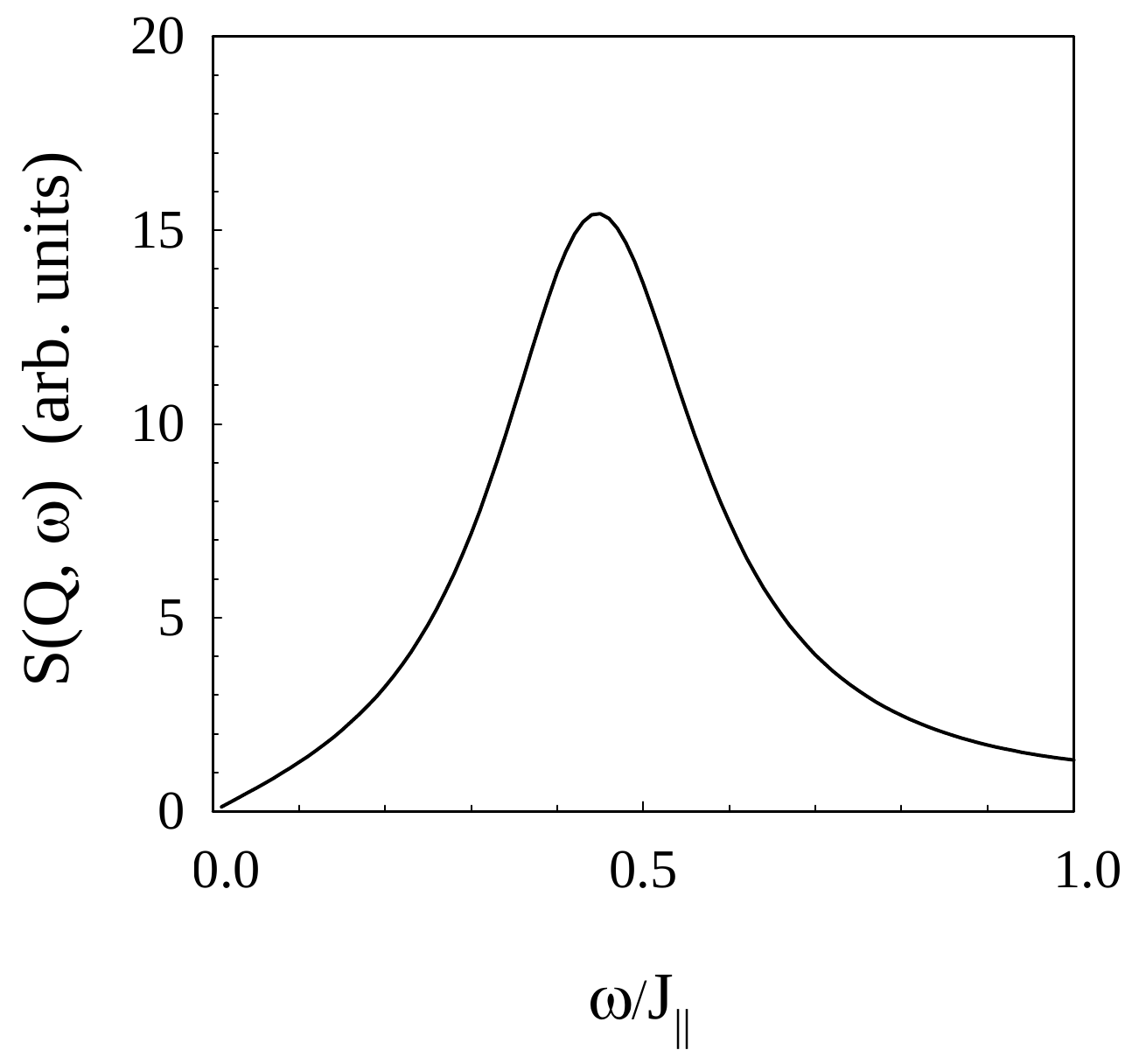} \caption{The resonance
energy $\omega_{r}$ at $p=0.20$ with $T=0$ for
$t_{\parallel}/J_{\parallel}=2.5$ and
$t_{\perp}/t_{\parallel}=0.72$. \label{fig4}}
\end{figure}

We are now ready to discuss the dynamical spin response of
Sr$_{14-x}$Ca$_{x}$Cu$_{24}$O$_{41}$ in the pressure-induced SC
state. The dynamical spin structure factor $S(k,\omega)$ in
the ($k,\omega$) plane at the doping concentration $p=0.20$ with
temperature $T=0$ for parameters $t_{\parallel}/J_{\parallel}=2.5$
and (a) $t_{\perp}/t_{\parallel}=0.4$ (underpressure) and (b)
$t_{\perp}/t_{\parallel}=0.8$ (overpressure) is plotted in Fig.
\ref{fig2} (hereafter we use the unit of $[2\pi]$). Obviously, the
mostly remarkable feature is the presence of an
incommensurate-commensurate transition in the spin fluctuation
geometry, where the magnetic excitation disperses with interchain
coupling (then pressure). In particular, the incommensurate spin
correlation in the pressure-induced SC state appears in the
underpressure regime, while the commensurate spin fluctuation
emerges in the overpressure regime. To check this point explicitly,
we plot the evolution of the magnetic scattering peaks at $p=0.20$
in $T=0$ for $t_{\parallel}/ J_{\parallel}=2.5$ with interchain
coupling (then pressure) for $\omega=0.4J_{\parallel}$ in Fig.
\ref{fig3}, where there is a {\it critical} value (then {\it
critical} pressure) of $t_{\perp}/t_{\parallel}\approx 0.72=P_{c}$,
which separates the pressure region into the underpressure
($t_{\perp}/t_{\parallel}<0.72$) and overpressure ($t_{\perp}/
t_{\parallel}> 0.72$) regimes, while $t_{\perp}/t_{\parallel}\approx
0.72$ is corresponding to the optimal pressure. In the underpressure
regime $t_{\perp}/t_{\parallel}<0.72$, the magnetic scattering peak
is split into two peaks at $[1/2\pm\delta]$ with $\delta$ as the
incommensurate parameter, and is symmetric around $[1/2]$. In this
case, spins are more likely to move along the legs of the ladders,
rendering the materials quasi-one-dimension. However, the range of
the incommensurate spin correlation decreases with increasing the
strength of the interchain coupling (then pressure), and then a
broad commensurate scattering peak appears in the optimal pressure
and overpressure regimes $t_{\perp}/ t_{\parallel}\geq 0.72$. In
particular, the magnetic resonance energy is located among this broad
commensurate scattering range. For determining this commensurate
magnetic resonance energy in the optimal pressure, we have made a
series of calculations for the intensities of the dynamical spin
structure factor $S(k,\omega)$ at $p=0.20$ with $T=0$ for
$t_{\parallel}/J_{\parallel}=2.5$ and
$t_{\perp}/t_{\parallel}=0.72$, and the result of the intensities
of $S(k,\omega)$ as a function of energy is plotted in Fig.
\ref{fig4}, where a commensurate resonance peak centered at
$\omega_{r}=0.45J_{\parallel}$ is obtained. Using an reasonably
estimative value \cite{Katano99} of $J_{\parallel}\sim 90$meV for
Sr$_{14-x}$Ca$_{x}$Cu$_{24}$O$_{41}$, the present result of the
resonance energy in the optimal pressure is $\omega_{r}\approx
40.5$meV. This anticipated spin gap $\Delta_{S}\approx 40.5$meV is
qualitatively consistent with the spin gap $\approx 32.5$meV
observed experimentally \cite{Katano99} on
Sr$_{14-x}$Ca$_{x}$Cu$_{24}$O$_{41}$.

\begin{figure}[h!]
\center\includegraphics[scale=0.5]{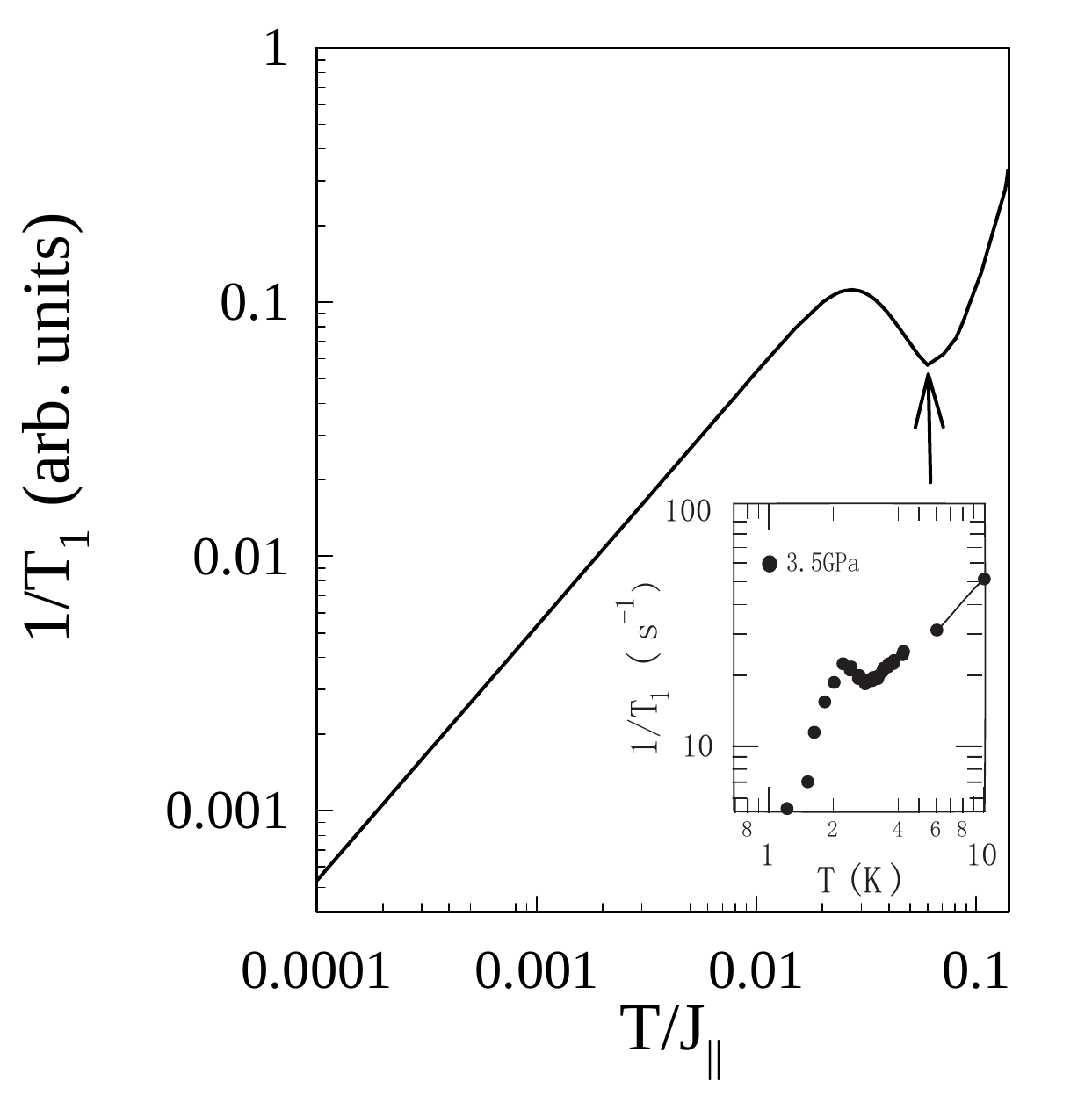} \caption{The
temperature dependence of the spin-lattice relaxation time $1/T_{1}$
in both logarithmic scales at $p=0.20$ for
$t_{\parallel}/J_{\parallel}=2.5$ and $t_{\perp}/t_{\parallel}=0.7$.
The arrow marks the position of the superconducting transition
temperature $T_{c}$. Inset: the experimental result on
Sr$_{14-x}$Ca$_{x}$Cu$_{24}$O$_{41}$ taken from Ref.
\cite{Fujiwara03}. \label{fig5}}
\end{figure}

In the dynamical spin response of the pressure-induced two-leg
ladder cuprate superconductors, one of the characteristic features
is the spin-lattice relaxation time $T_{1}$, which is closely
related to the dynamical spin structure factor (\ref{DSSF}), and can
be expressed as,
\begin{eqnarray}\label{T1}
{1\over T_{1}}={2K_{B}T\over g^{2}\mu_{B}^{2}\hbar}
\lim_{\omega\rightarrow 0}{1\over N}\sum_{k}F_{\alpha}^{2}(k)
{\chi''(k,\omega)\over\omega},
\end{eqnarray}
where $g$ is the lande-factor, $\mu_{B}$ is the Bohr magneton, and
$F_{\alpha}(k)$ is the form factors, while the dynamical spin
susceptibility $\chi''(k,\omega)=(1-e^{-\beta\omega})S(k,\omega)$.
Although the form factors $F_{\alpha}(k)$ have
dimension of energy, and the magnitude determined by atomic physics,
and the momentum dependence determined by geometry, however, for the
convenience, this form factors $F_{\alpha}(k)$ can be set to
constant without loss of generality \cite{He03}. In Fig. \ref{fig5},
we plot the spin-lattice relaxation time $1/T_{1}$ as a function of
temperature in both logarithmic scales at $p=0.20$ for
$t_{\parallel}/J_{\parallel}=2.5$ and $t_{\perp}/t_{\parallel}=0.7$
(underpressure), where we have chosen units $\hbar=K_{B}=1$. For
comparison, the corresponding experimental result \cite{Fujiwara03}
of Sr$_{14-x}$Ca$_{x}$Cu$_{24}$O$_{41}$ at $p\approx 0.20$ is also
shown in Fig. \ref{fig5}. The spin-lattice relaxation time $T^{-1}_{1}$
shows a linear temperature dependent behavior at low temperatures
($T>T_{c}$) followed passes through a minimum and displays a tendency
towards an increase with decreasing temperatures. In particular, it is
dominated by a peak developed below the SC transition temperature $T_{c}$.
Furthermore, this clear peak in $T^{-1}_{1}$ also confirms that a finite
SC gap exists in the quasiparticle excitation, then the spin-lattice
relaxation time under the SC transition temperature decreases
with decreasing temperatures, in qualitative agreement
with the experimental observation on Sr$_{14-x}$Ca$_{x}$Cu$_{24}$O$_{41}$
\cite{Fujiwara03}. In this case, this peak can be assigned to a SC
coherence peak while the temperature linear dependence of $T^{-1}_{1}$
at low temperatures to Korringa-type behavior. It is well-known that in
the conventional metals, the temperature-linear component in $T^{-1}_{1}$
in the normal state arises from paramagnetic free electrons
\cite{Fujiwara03}. However, in the present two-leg ladder cuprate
superconductors, the interaction between charge carriers and spins from
the kinetic energy term in the $t$-$J$ ladder (\ref{t-jmode1}) induces
the charge carrier-spin bound state in the normal state \cite{He03}. At
low temperatures ($T>T_{c}$), although the most of spins in the system
form the spin liquid state, the spin in the charge carrier-spin bound
state moves almost freely and therefore contributes to the
temperature-linear component in $T^{-1}_{1}$ \cite{He03}.

The essential physics of the pressure dependence of the dynamical
spin response in Sr$_{14-x}$Ca$_{x}$Cu$_{24}$O$_{41}$ in the
pressure-induced SC state is almost the same as in the normal state
case \cite{He03}. In the renormalized spin excitation spectrum
$\Omega^{2}_{k}=\omega^{2}_{1k}+B_{1k}{\rm Re}\Sigma^{(1)}_{s}
(k,\Omega_{k})$ in Eq. (\ref{DSSF}),
since both MF spin excitation spectrum $\omega_{1k}$ and spin
self-energy function $\Sigma^{(1)}_{s}(k,\omega)$ are strong
interchain coupling (then pressure) dependent, this leads to that
the renormalized spin excitation spectrum $\Omega_{k}$ also is
strong pressure dependent. Furthermore, the dynamical spin structure
factor in Eq. (\ref{DSSF}) has a well-defined resonance character,
where $S(k,\omega)$ exhibits peaks when the incoming neutron
energy $\omega$ is equal to the renormalized spin excitation, i.e.,
$W(k_{c},\omega)\equiv [\omega^{2}-\omega^{2}_{1k_{c}}-
B_{1k_{c}}{\rm Re}\Sigma^{(1)}_{s}(k_{c},\omega)]^{2}=
[\omega^{2}-\Omega_{k_{c}}^{2}]^{2}\sim 0$ for certain
critical wave vectors $k_{c}=k^{({\rm u})}_{c}$ in the
underpressure regime and $k_{c}=k^{({\rm o})}_{c}$ in
the optimal pressure and overpressure regimes, then the weight of
these peaks is dominated by the inverse of the imaginary part of the
spin self-energy $1/{\rm Im}\Sigma^{(s)}(k^{({\rm u})}_{c},\omega)$
in the underpressure regime and $1/{\rm Im}
\Sigma^{(s)}(k^{({\rm o})}_{c},\omega)$ in the optimal
pressure and overpressure regimes, respectively. In particular, for
the present spin self-energy ${\rm Re}\Sigma_{s}^{(1)}(k,\omega)=
{\rm Re}\Sigma_{L}^{(s)}(k,\omega)+{\rm Re}\Sigma_{T}^{(s)}(k,\omega)$,
${\rm Re}\Sigma_{L}^{(s)}(k,\omega)<0$ favors the one-dimensional behaviors,
while ${\rm Re}\Sigma_{T}^{(s)}(k,\omega)>0$ characterizes the quantum
interference between the chains in the ladders, therefore there is a
competition between ${\rm Re}\Sigma_{L}^{(s)}(k,\omega)$ and
${\rm Re}\Sigma_{T}^{(s)}(k,\omega)$. In the underpressure
regime, the main contribution for ${\rm Re}\Sigma_{s}^{(1)}(k,\omega)$
may come from ${\rm Re}\Sigma_{L}^{(s)}(k,\omega)$,
and spins and charge carriers are more likely to move along the
legs, then the incommensurate spin correlation emerges, where the
essential physics is almost the same as in the two-dimensional
$t$-$J$ model \cite{feng06}. Within the CSS fermion-spin framework,
as a result of self-consistent motion of charge carriers and spins,
the incommensurate spin correlation is developed, which means that
in the underpressure regime, the spin excitations drift away from
the AF wave vector, where the physics is dominated by the spin
self-energy ${\rm Re}\Sigma_{L}^{(s)}(k,\omega)$
renormalization due to charge carriers. However, the quantum
interference effect between the chains manifests itself by the
interchain coupling (then pressure), {\it i.e.}, this quantum
interference increases with increasing pressure. Thus in the optimal
pressure and overpressure regimes, ${\rm Re} \Sigma_{T}^{(s)}(k,\omega)$
may cancel the most incommensurate spin correlation
contributions from ${\rm Re} \Sigma_{L}^{(s)}(k,\omega)$, then
the commensurate spin fluctuation appears. In this sense, the
pressure is a crucial role to determine the symmetry of the spin
fluctuation in the two-leg ladder cuprate superconductors in the
pressure-induced SC state.

In summary, we have shown very clearly in this paper that if the
pressure effect is imitated by a variation of the interchain
coupling in the framework of the kinetic energy driven SC mechanism,
the dynamical spin structure factor of the $t$-$J$ ladder model
calculated in terms of the collective modes in the charge carrier
particle-hole and particle-particle channels per se can correctly
reproduce some main features found in the NMR and NQR measurements
on Sr$_{14-x}$Ca$_{x}$Cu$_{24}$O$_{41}$ in the pressure-induced SC
state, including the temperature dependence of the spin-lattice
relaxation time, without using adjustable parameters. The theory
also predicts that in the underpressure regime, the incommensurate
spin correlation appears, while the commensurate spin fluctuation
emerges in the optimal pressure and overpressure regimes, which
should be verified by further experiments.

\acknowledgments
JQ is supported by the National Natural Science
Foundation of China (NSFC) under Grant No. 11004006, YL is supported
by NSFC under Grant No. 11004084, and SF is supported by NSFC under
Grant No. 11074023, and the funds from the Ministry of Science and
Technology of China under Grant No. 2011CB921700.


\begin{thebibliography}{00}

\bibitem {Uehara96} Uehara Masatomo, Nagata Takashi, Akimitsu Jun,
Takahashi Hiroki, M\^ori Nobuo and Kinoshita Kyoichi 1996 \emph{J.
Phys. Soc. Jpn.} {\bf 65} 2764

\bibitem {Hiroi91} Hiroi Z, Azuma M, Takano M and Bando Y 1991 {\it J. Solid
State Chem.} {\bf 95} 230

\bibitem {Osafune97} Osafune T, Motoyama N, Eisaki H and Uchida S 1997
\emph{Phys. Rev. Lett.} {\bf 78} 1980

\bibitem {Eccleston98} Eccleston Roger S, Uehara Masatomo, Akimitsu Jun,
Eisaki Hiroshi, Motoyama Naoki and Uchida Shin-ichi 1998 \emph{Phys.
Rev. Lett.} {\bf 81} 1702

\bibitem{Katano99} Katano S, Nagata T, Akimitsu J, Nishi M and Kakurai K
1999 \emph{Phys. Rev. Lett.} {\bf 82} 636

\bibitem{Isobe98} Isobe M, Ohta T, Onoda M, Izumi F, Nakano S, Li J Q,
Matsui Y, Takayama-Muromachi E, Matsumoto T and Hayakawa H 1998
\emph{Phys. Rev.} B {\bf 57} 613

\bibitem{Dagotto99} See, e.g., the review, Dagotto E 1999 {\it Rep. Prog.
Phys.} {\bf 62} 1525, and references therein

\bibitem{Fujiwara03} Fujiwara Naoki, M\^ori Nobuo, Uwatoko Yoshiya,
Matsumoto Takehiko, Motoyama Naoki and Uchida Shinichi 2003
\emph{Phys. Rev. Lett.} {\bf 90} 137001; Fujiwara N, Fujimaki Y,
Uchida S, Matsubayashi K, Matsumoto T and Uwatoko Y 2009 \emph{Phys.
Rev.} B {\bf 80} 100503(R)

\bibitem{Piskunov04} Piskunov Y, J\'erome D, Auban-Senzier P, Wzietek P
and Yakubovsky A 2004 \emph{Phys. Rev.} B {\bf 69} 014510

\bibitem{Piskunov05} Piskunov Y, J\'erome D, Auban-Senzier P, Wzietek P
and Yakubovsky A 2005 \emph{Phys. Rev.} B {\bf 72} 064512

\bibitem {Anderson87} Anderson P W 1987 {\it Science} {\bf 235} 1196

\bibitem {He03} He Jianhui, Feng Shiping and Chen Wei Yeu 2003 \emph{Phys. Rev.} B
{\bf 67} 094402

\bibitem{feng04} Feng Shiping, Qin Jihong and Ma Tianxing 2004 {\it J.
Phys.: Condens. Matter} {\bf 16} 343

\bibitem{feng07} See, e.g., the review, Feng Shiping, Guo Huaiming,
Lan Yu and Cheng Li 2008 {\it Int. J. Mod. Phys.} B {\bf 22} 3757

\bibitem{Magishi98} Magishi K, Matsumoto S, Kitaoka Y, Ishida K, Asayama K,
Uehara M, Nagata T and Akimitsu J 1998 \emph{Phys. Rev.} B {\bf 57}
11533; Ohsugi S, Magishi K, Matsumoto S, Kitaoka Y, Nagata T and
Akimitsu J 1999 \emph{Phys. Rev. Lett.} {\bf 82} 4715

\bibitem{feng0306} Feng Shiping 2003 \emph{Phys. Rev.} B {\bf 68} 184501; Feng Shiping,
Ma Tianxing and Guo Huaiming 2006 {\it Physica} C {\bf 436} 14

\bibitem{Qin07} Qin Jihong, Chen Ting and Feng Shiping 2007 \emph{Phys. Lett.} A
{\bf 366} 611

\bibitem{Dagotto96} See, e.g., the review, Dagotto E and Rice T M 1996
{\it Science} {\bf 271} 618, and references therein

\bibitem {laughlin97} Laughlin R B 1997 \emph{Phys. Rev. Lett.} {\bf 79} 1726;
Laughlin R B 1995 {\it J. Low. Tem. Phys.} {\bf 99} 443

\bibitem {feng94} Feng Shiping, Su Z B and Yu L 1994 \emph{Phys. Rev.} B {\bf 49} 2368

\bibitem {plakida02} Plakida N M 2002 {\it Condens. Matter Phys.} {\bf 5} 707

\bibitem{Nagata98} Nagata T, Uehara M, Goto J, Akimitsu J, Motoyama N,
Eisaki H, Uchida S, Takahashi H, Nakanishi T and M\^ori N 1998
\emph{Phys. Rev. Lett.} {\bf 81} 1090

\bibitem{Schrieffer64} See, e.g., Schrieffer J R \emph{Theory of
Superconductivity} (Addison-Wesley, San Francisco, 1964)

\bibitem{Ohta97} Ohta Tomoko, Izumi Fujio, Onoda Mitsuko, Isobe Masaaki,
Takayama-Muromachi Eiji and Hewat Alan W 1997 \emph{J. Phys. Soc.
Jpn.} {\bf 66} 3107

\bibitem{Kato96} Kato Masatsune, Shiota Kazunori and Koike Yoji 1996
{\it Physica} C {\bf 258} 284

\bibitem{Qin06} Qin Jihong, Yuan Feng and Feng Shiping 2006 \emph{Phys. Lett.}
A {\bf 358} 448

\bibitem{qin02} Qin Jihong, Song Yun, Feng Shiping and Chen Wei Yeu 2002 \emph{Phys. Rev.}
B {\bf 65} 155117

\bibitem{feng06} Feng Shiping, Ma Tianxing and Wu Xintian 2006 \emph{Phys. Lett.}
A {\bf 352} 438

\end{thebibliography}
\end{document}